\documentclass[12pt,letterpaper]{article}
\usepackage{epsfig,rotating,setspace,latexsym,amsmath,epsf,amssymb,amsfonts,bm,theorem,cite,caption,subcaption,enumerate,longtable,accents}
\usepackage{algorithm,algorithmic,graphicx,epsf,authblk,epstopdf,url,color,multirow}
\usepackage{mathtools,comment,tabularx}
\usepackage{comment}
\usepackage{physics}

\usepackage{graphicx}
\usepackage{caption}
\usepackage{subcaption}
\usepackage[toc,title]{appendix}

\newtheorem{theorem}{Theorem}

\newtheorem{lemma}{Lemma}

\newenvironment{Proof}[1]{\medskip\par\noindent{\bf Proof:\,}\,#1}{{\mbox{\,$\blacksquare$}\par}}

\newcommand{\st}{{\text{s.t.}}}

\newcolumntype{Y}{>{\centering\arraybackslash}X}

\allowdisplaybreaks

\title{The Role of Early Sampling in Age of Information Minimization in the Presence of ACK Delays}

\author{Sahan Liyanaarachchi \qquad Sennur Ulukus\\
        \normalsize Department of Electrical and Computer Engineering\\
        \normalsize University of Maryland, College Park, MD 20742\\
        \normalsize  \emph{sahanl@umd.edu} \qquad \emph{ulukus@umd.edu}}

\begin{document}

\date{}

\maketitle

\vspace*{-1.0cm}

\begin{abstract}
    We study the structure of the optimal sampling policy to minimize the average age of information when the channel state (i.e., busy or idle) is not immediately perceived by the transmitter upon the delivery of a sample due to random delays in the feedback (ACK) channel. In this setting, we show that it is not always optimal to  wait for ACKs before sampling, and thus, \emph{early sampling} before the arrival of an ACK may be optimal. We show that, under certain conditions on the  distribution of the ACK delays, the optimal policy is a mixture of two threshold policies.
\end{abstract}

\section{Introduction} \label{sec:intro}
Sampling for data freshness has been an increasingly important problem due to its wide use cases in the wireless domain. Data freshness is often measured through a non-decreasing function of age of information (AoI), simplest being the instantaneous age of the process itself given by $\Delta_t=t-u(t)$, where $u(t)$ is the generation time of the freshest sample obtained from the observed process \cite{yates2020age}. Many of the previous work in this area involves modelling the communication system  as an \emph{enqueue-and-forward} model\cite{rts2012, rtsms2012, rtsms2019}, where the updates are generated randomly and enqueued before being transmitted to the receiver. However, recent works involve the \emph{generate-at-will} model introduced in \cite{lts2015} where the sampler has the ability to generate a sample when needed. In \cite{ys17}, this model has been studied for general age penalty functions, where it is shown that the \emph{zero-wait} policy is not always optimal. Most of the existing communication models consist of a single channel with a  transmission delay or erasures, and assume instantaneous feedback about channel state\cite{ys19, tme2022, tmef2023}. However, in a practical communication system, the channel carrying the feedback/ACK is non-ideal. The work in \cite{twd2020, twdsd2021} introduces a two-channel model, with a forward channel and a backward channel, to address this problem. This has been further extended in \cite{urtwd2022} by introducing an unreliable communication channel with packet drops. In all these models, it is assumed that the next sample should always be taken after receiving the ACK of the previous sample. Our paper extends this line of work by considering the possibility of \emph{early sampling}, where new samples may be generated before ACKs of previous samples are received, as needed. 

\begin{figure}[t]
    \centering
    \includegraphics[scale=1.25]{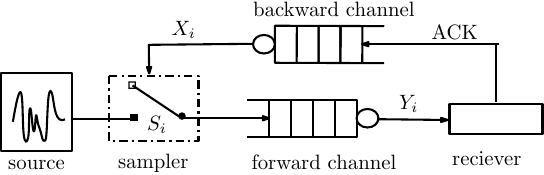}
    \caption{System model.}
    \label{fig:1}
\end{figure}

Consider  a two-channel communication model as shown in Fig.~\ref{fig:1}, where a transmitter observes a stochastic source and transmits the samples over a channel with a random delay (forward channel). Once a sample arrives at the receiver, it generates an acknowledgement message (ACK) which is sent to the transmitter again via a channel with a random delay (backward channel). The transmitter perceives the channel state through these ACKs. If these ACKs arrived at the transmitter instantaneously as they were generated, then the transmitter would always know the exact channel state of the forward channel at any given time. Under such circumstances, an optimal sampling policy should not generate a new sample when the channel is busy \cite{ys17}. If there is a delay in ACKs, the forward channel could become free at a time much earlier than the time at which the transmitter perceives it to be free. In this scenario, a naive approach would be to always wait for the ACK of the previous sample before sampling the next. In this work, we explore how to exploit the time window between knowing that the channel is free and the time at which the channel is actually free, by allowing the transmitter to sample before the arrival of ACKs of the previous samples. 

Here, we consider a \textit{generate-at-will} model with preemptive transmissions \cite{preempt1,preempt2,preempt3},  which enables the transmitter to take a sample and transmit it at any time. Since we allow sampling before ACKs, the following questions must be addressed first:

\begin{itemize}
    \item \textbf{What does it mean to transmit when the channel is busy?} We model the forward channel as a queue with possible preemption. If a sample is generated and attempted to be transmitted when the channel is busy, we assume that this new sample gets corrupted during its storage into the queue. However, this corrupted sample does not affect the transmission of the sample that is being transmitted. As the queue passively serves what is stored, once the current sample has finished its transmission, the next sample (corrupted) in the queue  will be served unless a preemptive transmission is initiated by the transmitter.
    
    \item \textbf{How are the ACKs generated at the receiver?} When a new sample is received, an ACK is generated which contains the delivery time of the sample. If the new sample is received while sending back the ACK of an old sample, then as was in the transmitter side, we consider that the newly generated ACK will be corrupted. Under certain conditions on the distribution of the transmission delay and the ACK delay, the collision in ACKs can be eliminated. These conditions will be discussed in the next section and are assumed to hold throughout the paper.

    \item \textbf{When are preemptive transmissions initiated?} If a corrupted sample gets transmitted by the forward channel, it would not reduce the age of the process and therefore would have wasted valuable transmission time on a corrupted sample. If the transmitter knows that a corrupted sample is being transmitted, then it is always better to cancel the current transmission and transmit an uncorrupted sample if possible. On the other hand, it is not always ideal to cancel the transmission of an uncorrupted sample. Therefore, we assume that the transmitter would only initiate a preemptive transmission if the transmitter is certain that a corrupted sample is being transmitted when we take the new sample. We further assume that, if a preemptive transmission is initiated, then all samples (corrupted) in the queue would be dropped.
    
    \begin{figure}[t]
    \centering
    \begin{subfigure}{\textwidth}
        \centering
        \includegraphics{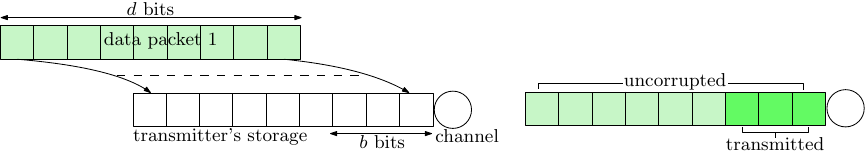}
        \caption{Arrival of a data packet when the storage is empty.}
    \end{subfigure}
    \par\bigskip
    \begin{subfigure}{\textwidth}
        \centering
        \includegraphics{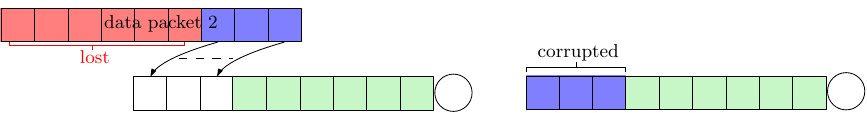}
        \caption{Arrival of a data packet when the storage is non-empty.}
    \end{subfigure}
    \caption{Physical model.}
    \label{fig:phy}
    \end{figure}

    \item \textbf{Why are enqueued samples assumed to be corrupted?} In the actual physical model (see Fig.~\ref{fig:phy}), we assume that the data packets involved are of a fixed size ($d$ bits) and the transmitter is only capable of accommodating (storing) $d$ bits at any given time. Suppose these $d$ bits are initially empty. When a sample is taken, it will be written on to these $d$ bits and these $d$ bits will be sequentially transmitted across the channel (say $b$ bits at a time) where the total time to transmit these $d$ bits would correspond to the channel service time. Once $b$ bits have been transmitted, $b$ bits from the transmitter's storage will be relieved. Once a new sample is taken, the bits of the new data packet is written onto the next available bits in the storage space of the transmitter. In doing so, some of the bits of this new sample would be lost and therefore we assume that the current sample is corrupted. Once the channel has finished serving the last $b$ bits of the initial data packet, since the next $b$ bits in the transmitter's storage is non-empty, it will start another cycle of transmission and therefore will start serving the corrupted sample until a total of $d$ bits have been transmitted. If the transmitter has the knowledge that a corrupted sample is being served, it can initiate a preemptive transmission by clearing the $d$ storage bits and storing a new sample in them. If a new data packet arrives when the storage is full, then that data packet is completely lost. This type of a physical model is common in small IoT devices which are often used in remote estimation settings. Therefore, we abstract this physical model with a queue where the queued up samples are considered to be corrupted with probability 1 if the queue is currently serving a sample. This queuing model is a variant of the erasure-queue channel which is commonly used in quantum communication models where stored qubits suffer from a waiting time dependent decoherence \cite{eqchannels}. The version of the problem where multiple samples may be saved in the queue and served sequentially over time is an interesting extension of the simpler model studied in this paper.
\end{itemize}

Under the above model assumptions, we show that the system model oscillates between two distinct states, where in state 1 we sample knowing that the channel is busy and in state 2 we sample knowing that the channel is free (or can be made free via preemption). We show that the structure of the optimal stationary deterministic sampling policy that minimizes the average age of information is a mix of two threshold policies, one for each state.

\section{Problem Formulation} \label{sec:formulation}
We say that a sample was correctly received, if it was not corrupted before the transmission by the forward channel. Let $S_0, S_1, \ldots$  denote the sequence of sampling times of correctly received samples where  $S_i\leq S_{i+1}$. Let the sequence of the  forward channel service times (transmission delays) and ACK delays be represented by $\{Y_i\overset{\mathrm{iid}}{\sim}Y\}_{i=0}^{\infty}$ and $\{X_i\overset{\mathrm{iid}}{\sim}X\}_{i=0}^{\infty}$, respectively. We assume that $Y$ and $X$ have finite first and second moments. Denote by $D_i$ the delivery time of the $i$th correctly received sample and by $N_i$ the total number of samples taken by the time $S_i$. Let $\pi$ be a causal stationary deterministic policy, $f_{max}$ be the maximum allowable sampling rate, and $\Delta_t$ be the instantaneous AoI of the samples at the receiver. Then, the problem of minimizing the average AoI can be expressed as follows, 
\begin{align} \label{eqn:1}
    \min_{\pi} & \quad \limsup_{n \to \infty}{\frac{\mathbb{E}\left[\int_{0}^{D_n} \Delta_t \,dt \right]}{\mathbb{E}[D_n]}} \nonumber\\
    \st & \quad \liminf_{n \to \infty}{\mathbb{E}\left[\frac{S_n}{N_n}\right]} \geq \frac{1}{f_{max}}
\end{align}

Solving for the optimal solution in problem (\ref{eqn:1}) can deem difficult for a general distribution of $X$ and $Y$ due to complex scenarios such as ACK collisions. Therefore, to simplify the problem, we assume that the distributions of ACK delays and forward channel service times satisfy the condition $X\leq Y$ almost surely (a.s.). This can be argued to be a reasonable assumption in many practical scenarios since in a general communication protocol, the packet size of ACKs is much smaller than data packets, and hence, would almost surely be received faster than the data packets. Under the above assumption, Lemma~\ref{lem:1} below uncovers an important structural property of the optimal policy.

\begin{figure} [t]
    \centering
    \includegraphics[width=\textwidth]{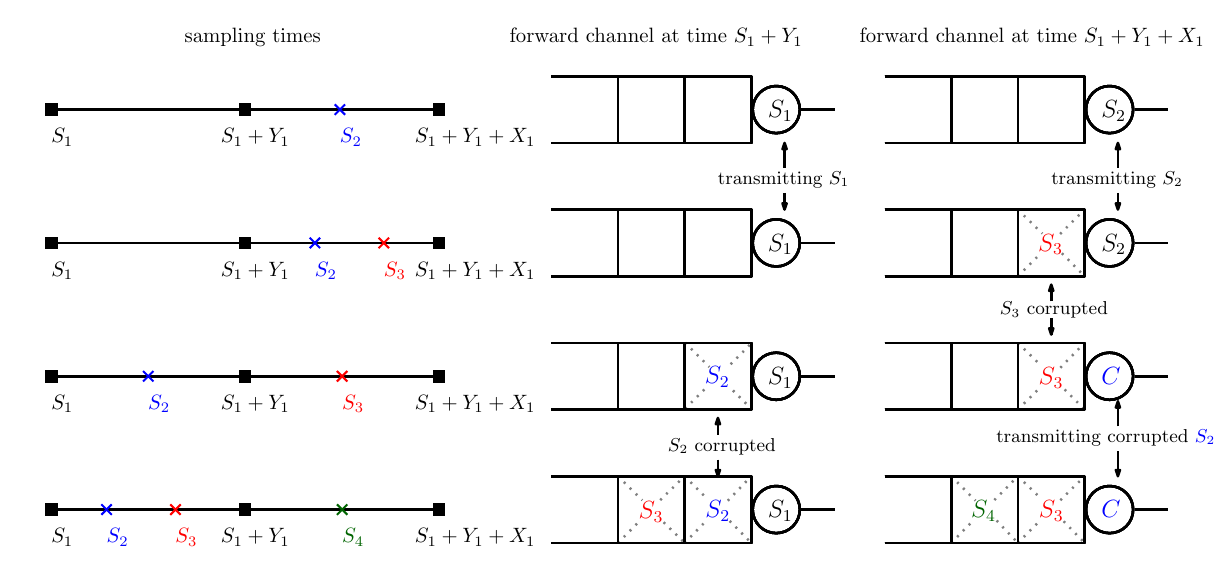}
    \caption{Transmitting before ACKs.}
    \label{fig:erasures}
\end{figure}

\begin{lemma} \label{lem:1}
    If $X\leq Y$ a.s., then under an optimal sampling policy, one should not take more than 1 sample before receiving the ACK of the previous sample.
\end{lemma}

\begin{Proof}
    Let  $S_i$ be the sampling time of a correctly received sample. Then, $D_i=S_i+Y_i$ is the time of its delivery and $A_i=S_i+Y_i+X_i$ is the time at which the transmitter receives its ACK. Suppose the transmitter takes another sample $\tilde{S}_{i+1}$ before the ACK of $S_i$ has arrived at the transmitter. If $\tilde{S}_{i+1}$ was taken before $D_i$, then $\tilde{S}_{i+1}$ would be corrupted and at $D_i$, this corrupted sample will start its transmission. Since $X\leq Y$ a.s., the delivery time of this corrupted sample would be after $A_i$. Therefore, the channel would be busy in the time interval  $(S_i,A_i)$ and any sample taken after $\tilde{S}_{i+1}$ until $A_i$ would be corrupted again in the forward channel. If $\tilde{S}_{i+1}$ was taken after $D_i$, it will not be corrupted, and since $X\leq Y$ a.s., $\tilde{S}_{i+1}$ would again be delivered only after receiving the ACK of $S_i$. Thus, the channel will be busy. Hence, any samples taken in the interval $(\tilde{S}_{i+1},A_{i})$ would again be corrupted in the forward channel. No preemptive transmissions will take place in the interval $(S_i,A_i)$ as we would not know the exact status of the sample that is being currently served by the forward channel until we get the ACK of $S_i$. Hence, it is not optimal to take more than 1 sample before an ACK arrives. The other samples should be taken after receiving the ACK.
\end{Proof}

\begin{figure}[t]
    \centering
    \includegraphics{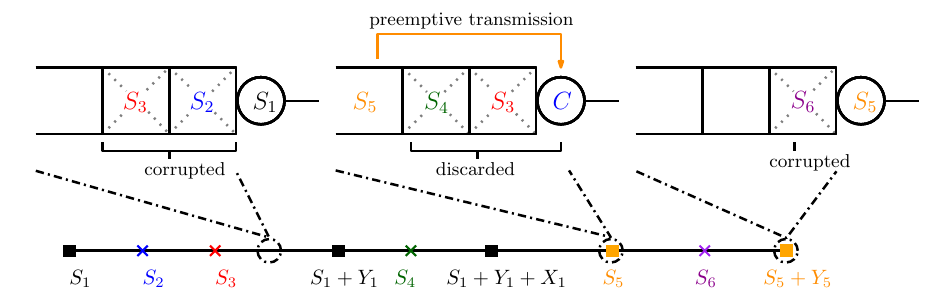}
    \caption{Preemptive transmission.}
    \label{fig:preempt}
\end{figure}

Under the assumption that $X\leq Y$ a.s., Lemma~\ref{lem:1} shows that only at most one sample may be taken before receiving the ACK of the previous sample. Moreover, the delivery time of this sample (either corrupted or not) would fall after the time of reception of the ACK of the previous sample. Therefore, there will be no collision in ACKs. Following the same nomenclature in Lemma~\ref{lem:1}, let  $S_i$, $D_i$ and $A_i$ be the sampling time, delivery time and the acknowledgement time of a correctly received sample. Let $\tilde{S}_{i+1}$ be the sampling time of the next sample. Since we send back the delivery time $D_i$ along with the ACK, if  the next sample was taken at a time $\tilde{S}_{i+1}<A_i$, then at time $A_i$ we exactly know if the new sample was corrupted or not. When we receive the ACK at time $A_i$, if $\tilde{S}_{i+1}<D_i$, we know the new sample got corrupted, and therefore, the channel is serving a corrupted sample. If we know the channel is serving a corrupted sample, we can free up the channel through a preemptive transmission of the next sample. At $A_i$, if  $D_i\leq \tilde{S}_{i+1}< A_i$, we know the new sample will be successfully transmitted, and therefore, the channel is busy serving an uncorrupted sample. If $ A_i<\tilde{S}_{i+1}$, then we know the channel is definitely free. 

\begin{figure}[t]
    \centering
    \includegraphics[width=\textwidth]{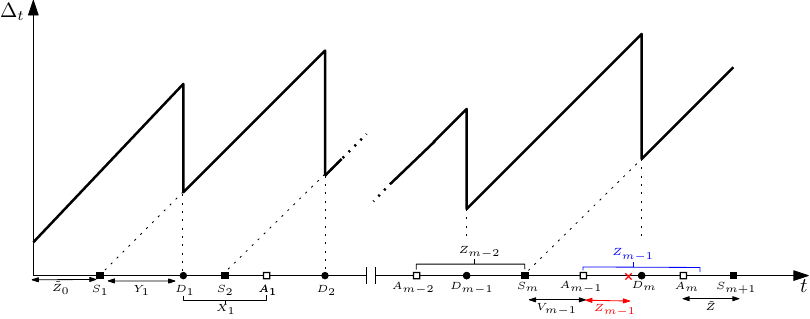}
    \caption{A typical transition cycle.}
    \label{fig:cycle}
\end{figure}

Therefore, we can characterize the system into two states based on the information available to the transmitter when an ACK arrives. In state 1, we  have the knowledge that the channel is busy serving an uncorrupted sample and in state 2 we have the knowledge that the channel is free (or can be made free via preemption). If the system is in state 1, when we receive an ACK if we had already taken the next sample and know it was corrupted (depicted by red $Z_{m-1}$ in Fig.~\ref{fig:cycle}) or if we have not taken the next sample by the time we received the ACK (depicted by blue $Z_{m-1}$ in Fig.~\ref{fig:cycle}), the system would make a transition from state 1 to state 2. Otherwise it would stay in state 1. If the system is in state 2 and we take a sample then it would directly revert back to state 1. Thus, the system model would consist of cycles of multiple state 1 to state 1 transitions, followed by a state 1 to state 2 to state 1 transition. Fig.~\ref{fig:cycle} shows one such transition cycle. 

After receiving an ACK, let $Z$ be the waiting time  before taking the next sample when in state 1 and $\tilde{Z}$ be the waiting time before taking the next sample when in state 2. Any sampling policy under consideration can be characterized using the waiting times in the two system states. Therefore, the goal of this paper is to find these waiting times based on the system state, previous transmission times and delivery times available to the system when an ACK arrives. Since we are only considering the stationary deterministic policies, the problem (\ref{eqn:1}) reduces to determining the optimal waiting times for one system state transition cycle. A stationary policy in this setting is defined as a policy which induces a stationary distribution among these two system states. 

Let $\tau$ be the time duration of one transmission cycle and $N_{\tau}$ be the number of samples taken in that transmission cycle. Then, the problem (\ref{eqn:1}) can be expressed as,
\begin{align} \label{eq:2}
    \min_{\pi} & \quad \frac{\mathbb{E}\left[\int_{0}^{\tau} \Delta_t \,dt \right]}{\mathbb{E}[\tau]} \nonumber\\
    \st & \quad {\frac{\mathbb{E}[\tau]}{\mathbb{E}[N_{\tau}]}} \geq \frac{1}{f_{max}}
\end{align}

We say that a policy is optimal if it solves (\ref{eq:2}) exactly, and a policy is asymptotically optimal if the policy becomes optimal as $f_{max}$ goes to $\infty$. Next, we show that under an optimal policy, the waiting time in state 2 should be a function of the previous transmission delay and ACK delay. Further, under an asymptotically optimal policy, when in state 1, we should not wait more than a constant time period $K$ to sample before an ACK arrives and if an ACK comes before $K$ time units have elapsed from the previous sample, then the waiting time should be a function of the previous transmission delay and the ACK delay.

\begin{lemma}\label{lem:2}
     If $X\leq Y$ a.s.~and $\inf X +\inf Y \leq \sup Y$, then the optimal waiting time $\tilde{Z}$ in state 2 must be a function of the previous transmission time and the ACK delay. The  asymptotically optimal waiting time $Z$ in state 1 should be a function of $V$ which is the time elapsed from previous sample  to the time we received the ACK and further $Z(V)+V = K$, where $K$ is some constant.
\end{lemma}

\begin{Proof}
    Consider the transition cycle starting from the first time we transitioned to state 1 and denote this time by $S_1$. Let $M$ be the number of samples taken before transitioning to state 2. $M$ is a stopping time which depends on our waiting policy in state 1. For $i< M$, let $Z_{i-1}$ be the wait time to take the sample $S_{i+1}$ after receiving the ACK for $S_{i-1}$ at time $A_{i-1}$ and let $V_{i-1}= A_{i-1}-S_i$ with $A_0=S_1$. Therefore, trivially $V_0=0$, $S_{i+1} = A_{i-1}+Z_{i-1}$ and $A_{i-1} = S_i + V_{i-1}$. Let $\tilde{Z}$ be the wait time after reaching state 2 to take the next sample. Under the condition that $\inf X +\inf Y \leq \sup Y$, for any waiting policy employed in state 1, the probability of transition from state 1 to state 1 is always bounded below one. Therefore, even though the transition probabilities of the system states may vary with the observed delays and transmission times, since the state  space is finite and transition probabilities are always uniformly bounded below one, $M$ will have finite expectation, i.e., $\mathbb{E}[M]<\infty$. Theorem~\ref{thrm:3} handles the case for $\sup Y< \inf X +\inf Y$. Under the above assumptions, the the average AoI can be expressed as follows,
    \begin{align}
      &\frac{\mathbb{E}\left[\int_{0}^{\tau} \Delta_t \,dt \right]}{\mathbb{E}[\tau]}\nonumber \\
      &=\frac{\mathbb{E}\left[\int_{S_1}^{S_1+Y_1}(t-S_0) \,dt + \sum_{i=1}^{M-1} \int_{S_i+Y_i}^{S_{i+1}+Y_{i+1}}(t-S_i) \,dt + \int_{S_M+Y_M}^{S_M+Y_M+X_M+\tilde{Z}}(t-S_M) \,dt\right]}{\mathbb{E}\left[\sum_{i=0}^{M-2}(V_i+Z_i)+Y_M+X_M+\tilde{Z}\right]} \nonumber \\
      &=\frac{\mathbb{E}\left[\int_{S_0+Y_0+X_0+\tilde{Z}_0}^{S_0+Y_0+X_0+\tilde{Z}_0+Y_1}(t-S_0) \,dt + \sum_{i=1}^{M-1} \int_{S_i+Y_i}^{S_i+V_{i-1}+Z_{i-1}+Y_{i+1}}(t-S_i) \,dt + \int_{S_M+Y_M}^{S_M+Y_M+X_M+\tilde{Z}}(t-S_M) \,dt\right]}{\mathbb{E}\left[\sum_{i=0}^{M-2}(V_i+Z_i)+Y_M+X_M+\tilde{Z}\right]} \nonumber\\
      &=\frac{\mathbb{E}\left[\int_{Y_0+X_0+\tilde{Z}_0}^{Y_0+X_0+\tilde{Z}_0+Y_1}t \,dt + \sum_{i=1}^{M-1} \int_{Y_i}^{V_{i-1}+Z_{i-1}+Y_{i+1}}t \,dt + \int_{Y_M}^{Y_M+X_M+\tilde{Z}}t \,dt\right]}{\mathbb{E}\left[\sum_{i=0}^{M-2}(V_i+Z_i)+Y_M+X_M+\tilde{Z}\right]} \label{eq:3}
    \end{align}
In (\ref{eq:3}), $S_0$ is the last sample in the previous transition cycle, $Y_0, X_0, \tilde{Z_0}$ have the same distribution as $Y_M,X_M,\tilde{Z}$ and the convention $\sum_{i=1}^0f(i)=\sum_{i=0}^{-1}f(i) = 0$ is assumed. The numerator can be further simplified by observing that,
\begin{align}\label{eqn:4}
    \mathbb{E}\left[\int_{Y_0+X_0+\tilde{Z}_0}^{Y_0+X_0+\tilde{Z}_0+Y_1}t \,dt +\int_{Y_M}^{Y_M+X_M+\tilde{Z}}t \,dt\right] =  \mathbb{E}\left[\int_{Y_M}^{Y_M+X_M+\tilde{Z}+Y_{M+1}}t \,dt \right] 
\end{align}
Thus, the optimization in problem (\ref{eq:2}) is equivalent to the following optimization problem,
\begin{align} \label{eqn:5}
   AoI_{opt} = \min_{Z(.),\tilde{Z}(.)} & \quad \frac{\mathbb{E}\left[\sum_{i=1}^{M-1} \int_{Y_i}^{V_{i-1}+Z_{i-1}+Y_{i+1}}t \,dt\right] + \mathbb{E}\left[\int_{Y_M}^{Y_M+X_M+\tilde{Z}+Y_{M+1}}t \,dt  \right]}{\mathbb{E}\left[\sum_{i=1}^{M-1}(V_{i-1}+Z_{i-1})+Y_M+X_M+\tilde{Z}\right]} \nonumber\\
    \st & \quad \frac{\mathbb{E}\left[\sum_{i=1}^{M-1}(V_{i-1}+Z_{i-1})+Y_M+X_M+\tilde{Z}\right]}{\mathbb{E}\left[M+\textbf{1}_e\right]} \geq \frac{1}{f_{max}}
\end{align}

In (\ref{eqn:5}), $\textbf{1}_e$ is the indicator of the event $\{Y_M>V_{M-1}+Z_{M-1}\}$ where the system transitions to state 2  because of a corrupted sample. Note that since $V_{i-1}+Z_{i-1}<Y_i+X_i$ for $i<M$, $\mathbb{E}[X]<\infty$, $\mathbb{E}[Y]<\infty$ and $\mathbb{E}[M]<\infty$, the following holds true: $\mathbb{E}\left[\sum_{i=1}^{M-1}(V_{i-1}+Z_{i-1})\right]\leq \mathbb{E}\left[\sum_{i=1}^{M-1}(Y_i+X_i)\right]=\mathbb{E}[M]\mathbb{E}[X+Y]<\infty$. The last equality is obtained from the Wald's identity. Similarly, we can show that  $ \mathbb{E}\left[ \sum_{i=1}^{M-1} \int_{Y_i}^{V_{i-1}+Z_{i-1}+Y_{i+1}}t \,dt \right]<\infty$ since $\mathbb{E}[Y^2]<\infty$ and $\mathbb{E}[X^2]<\infty$. In the same manner, it follows that $\mathbb{E}\left[Y_M+X_M\right]$ and $\mathbb{E}\left[(Y_M+X_M)^2\right]$ are finite as well for any given policy in state 1.

We give structure of an optimal policy in two steps. First, we will show that for any given waiting policy $Z$ employed in state 1, the optimal waiting policy $\tilde{Z}$ in state 2 is a function only dependent on $Y_M$ and $X_M$. Next, we show that for any given $\tilde{Z}$ which is a function of $Y_M$ and $X_M$, the asymptotically optimal waiting policy in state 1 should be such that $V_{i-1}+Z_{i-1}$ is a constant for all $i$. To prove this, we follow an approach similar to \cite{ys19}. Consider the following auxiliary optimization problem,
\begin{align}\label{eqn:6}
    J(c)= \min_{Z(.),\tilde{Z}(.)} & \quad \mathbb{E}\left[ \sum_{i=1}^{M-1} \int_{Y_i}^{V_{i-1}+Z_{i-1}+Y_{i+1}}t \,dt +\int_{Y_M}^{Y_M+X_M+\tilde{Z}+Y_{M+1}}t \,dt\right] \nonumber \\
    & \quad - c\mathbb{E}\left[\sum_{i=1}^{M-1}(V_{i-1}+Z_{i-1})+Y_M+X_M+\tilde{Z}\right] \nonumber\\
     \st & \quad \frac{\mathbb{E}\left[\sum_{i=1}^{M-1}(V_{i-1}+Z_{i-1})+Y_M+X_M+\tilde{Z}\right]}{\mathbb{E}\left[M+\textbf{1}_e\right]} \geq \frac{1}{f_{max}}
\end{align}
It can be easily shown that $c\lesseqgtr AoI_{opt} \iff J(c) \lesseqgtr 0$ and when $J(c)=0$, the optimal solutions are identical \cite{dinkelbach}. Therefore, the structure of the optimal solutions are the same when $c=AoI_{opt}$. Let us first consider the minimization with respect to $\tilde{Z}$ for a given policy $Z$. Note that given the policy $Z$, the above optimization problem is convex with respect to the functional $\tilde{Z}$ (see Appendix~\ref{apd:cvx} for a proof). Consider the following Lagrangian,
\begin{align} \label{eqn:7}
    L(\tilde{Z},\lambda)=& \mathbb{E}\left[\sum_{i=1}^{M-1}\left[\int_{Y_i}^{V_{i-1}+Z_{i-1}+Y_{i+1}}t \,dt-(c+\lambda)(V_{i-1}+Z_{i-1})\right]\right]\nonumber\\
    &+\mathbb{E}\left[\int_{Y_M}^{Y_M+X_M+\tilde{Z}+Y_{M+1}}t \,dt-(c+\lambda)(Y_M+X_M+\tilde{Z})\right] + \lambda\frac{\mathbb{E}\left[M+\textbf{1}_e\right]}{f_{max}}
\end{align}
For the dual problem of the Lagrangian with fixed $c+\lambda$, the term related to the control decision $\tilde{Z}$ is $\mathbb{E}\left[\int_{Y_M}^{Y_M+X_M+\tilde{Z}+Y_{M+1}}t \,dt-(c+\lambda)(Y_M+X_M+\tilde{Z})\right]$. $Y_{M+1}$ is independent of $Y_M$ and $X_M$. Therefore, given $Y_M$ and $X_M$, $(Y_M,X_M)$ is a sufficient statistic for determining $\tilde{Z}$. Thus, $\tilde{Z}$ must be a function of $(Y_M,X_M)$ in the minimization of the dual problem. For any given policy $Z$, we can always find a $\tilde{Z}$ such that the sampling constraint is strictly satisfied and hence strong duality applies for the Lagrangian $L(\tilde{Z},\lambda)$. Therefore, for a given policy $Z$, $\tilde{Z}$ must be a function of $(Y_M,X_M)$ in the original optimization problem as well.

Now, let us look at the minimization of the problem (\ref{eqn:6}) with respect to $Z$ given that $\tilde{Z}$ is a fixed function of $(Y_M,X_M)$. For that, let us look at the following Lagrangian,
\begin{align}
     L(Z,\lambda) &= \mathbb{E}\left[ \sum_{i=1}^{M-1} \int_{Y_i}^{V_{i-1}+Z_{i-1}+Y_{i+1}}t \,dt +\int_{Y_M}^{Y_M+X_M+\tilde{Z}+Y_{M+1}}t \,dt\right] \nonumber\\
     & \quad \quad - (c+\lambda)\mathbb{E}\left[\sum_{i=1}^{M-1}(V_{i-1}+Z_{i-1})+Y_M+X_M+\tilde{Z}\right] + \lambda\frac{\mathbb{E}\left[M+\textbf{1}_e\right]}{f_{max}} \\
     &= \mathbb{E}\Bigg[ \sum_{i=1}^{\infty} \left(\int_{Y_i}^{V_{i-1}+Z_{i-1}+Y_{i+1}}t \,dt\right)\textbf{1}_{\{M>i\}}+\left(\int_{Y_i}^{Y_i+X_i+\tilde{Z}+Y_{i+1}}t \,dt\right)\textbf{1}_{\{M=i\}}\Bigg] \nonumber\\
     &\quad \quad -(c+\lambda)\mathbb{E}\Bigg[ \sum_{i=1}^{\infty} \left(V_{i-1}+Z_{i-1}\right)\textbf{1}_{\{M>i\}}+(Y_i+X_i+\tilde{Z})\textbf{1}_{\{M=i\}}\Bigg]\nonumber\\
     & \quad \quad + \frac{\lambda}{f_{max}}\mathbb{E}\left[\sum_{i=1}^{\infty}\textbf{1}_{\{M>i\}} + \textbf{1}_{\{Y_i>V_{i-1}+Z_{i-1}\}}\textbf{1}_{\{M=i\}}\right] +\frac{\lambda}{f_{max}}\label{eqn:9}\\ 
     &= \sum_{i=1}^{\infty}\mathbb{E}\Bigg[\left(\int_{Y_i}^{Y_i+X_i+\tilde{Z}+Y_{i+1}}t \,dt-(c+\lambda)(Y_i+X_i+\tilde{Z})+\frac{\lambda}{f_{max}}\textbf{1}_{\{Y_i>V_{i-1}+Z_{i-1}\}}\right)\textbf{1}_{\{M=i\}} \nonumber\\
     & \quad \quad + \left(\int_{Y_i}^{V_{i-1}+Z_{i-1}+Y_{i+1}}t \,dt-(c+\lambda)(V_{i-1}+Z_{i-1}) +\frac{\lambda}{f_{max}}\right)\textbf{1}_{\{M>i\}} \Bigg] +\frac{\lambda}{f_{max}}\label{eqn:10}\\
     &=\sum_{i=1}^{\infty}\mathbb{E}\left[\mathbb{E}\left[\textbf{1}_{\{M>i-1\}}g(Y_i,X_i,V_{i-1},Z_{i-1},Y_{i+1})|I_i\right]\right] +\frac{\lambda}{f_{max}}\label{eqn:11}\\
     &=\sum_{i=1}^{\infty}\mathbb{E}\left[\textbf{1}_{\{M>i-1\}}\mathbb{E}\left[g(Y_i,X_i,V_{i-1},Z_{i-1},Y_{i+1})|I_i\right]\right] +\frac{\lambda}{f_{max}}\label{eqn:12}
\end{align}    
where $I_i=(V_{j-1},Y_{j-1},X_{j-1})_{1\leq j \leq i}$ is the information available at the transmitter when in state 1 and $g(Y_i,X_i,V_{i-1},Z_{i-1},Y_{i+1})$ is the term controlled by the control decision $Z_{i-1}$ which is given by,
\begin{align}\label{eqn:13}
    g(\cdot,\cdot,\cdot,\cdot,\cdot) &=\left(\int_{Y_i}^{Y_i+X_i+\tilde{Z}+Y_{i+1}}t \,dt-(c+\lambda)(Y_i+X_i+\tilde{Z}) +\frac{\lambda}{f_{max}}\textbf{1}_{\{Y_i>V_{i-1}+Z_{i-1}\}}\right)\textbf{1}_{\{B_i^c\}} \nonumber\\
     & \quad+  \left(\int_{Y_i}^{V_{i-1}+Z_{i-1}+Y_{i+1}}t \,dt-(c+\lambda)(V_{i-1}+Z_{i-1}) +\frac{\lambda}{f_{max}}\right)\textbf{1}_{\{B_i\}}
\end{align}
Here, $\textbf{1}_{\{B_i\}}$ is the indicator of the event, $\{Y_i<V_{i-1}+Z_{i-1}<Y_i+X_i\}$ and $\textbf{1}_{\{B_i^c\}}=1-\textbf{1}_{\{B_i\}}$, (\ref{eqn:9}) is obtained using the fact that $\mathbb{E}[M]<\infty$, and  (\ref{eqn:10}) is justified by applying Fubini Tonelli theorem for the individual expectations in (\ref{eqn:9}). Moving from (\ref{eqn:10}) to (\ref{eqn:11}), we have used the tower property of expectation along with the fact that $\textbf{1}_{\{M>i\}}=\textbf{1}_{\{M>i-1\}}*\textbf{1}_{\{B_i\}}$ and $\textbf{1}_{\{M=i\}}=\textbf{1}_{\{M>i-1\}}*\textbf{1}_{\{B_i^c\}}$. Then, (\ref{eqn:12}) follows immediately upon noticing that $\textbf{1}_{\{M>i-1\}}$ is completely deterministic given $I_i$.  Since $X<Y$ and $Y$ is independent of $X$, there exists a $u>0$ such that $X<u<Y$ a.s. Therefore, $V_{i-1}< X_{i-1} < u$. Since $u<Y_i$ regardless, the distribution of $Y_i$ is independent of $V_{i-1}$. Therefore, $Y_i$ and $Y_{i+1}$ are independent of $I_i$, and $V_{i-1}$ is a sufficient statistic for determining $Z_{i-1}$. Thus, the control decision $Z_{i-1}$ should be a function of $V_{i-1}$. Furthermore, $\mathbb{E}\left[g(Y_i,X_i,V_{i-1},Z_{i-1},Y_{i+1})|I_i\right]$ is controlled by  $Z_{i-1}$ through the term $V_{i-1}+Z_{i-1}$. Since  $(Y_i,X_i,Y_{i+1})$ is independent of $I_i$, the distribution of $(Y_i,X_i,Y_{i+1})$ is the same irrespective of $i$. Therefore, since we are essentially trying to minimize the same term at each stage in state 1, $V_{i-1}+Z_{i-1}$ must be the same constant for all $i$. Note that the constant that minimizes (\ref{eqn:13}) depends on $\lambda$. Since the problem is not necessarily convex with respect to the waiting times in state 1, the existence of an optimal $\lambda$ cannot be guaranteed (i.e., strong duality is not guaranteed) for all $f_{max}$. However, as $f_{max}$ goes to $\infty$, the sampling constraint would be inactive and therefore strong duality is guaranteed by setting $\lambda =0$. Hence, this structure of the waiting policy in state 1 is only \emph{asymptotically} optimal.
\end{Proof}

Lemma~\ref{lem:2} shows that when in state 1 we should sample at constant period $K$ until we transition to state 2, and when in state 2 we should sample only after waiting for a time period which is determined by the transmission and delivery time of the previous correctly received sample. These two structural properties can be used to further simplify the  problem (\ref{eq:2}) to a minimization problem that solves for the constant period $K$ and the waiting time function $\tilde{Z}(Y,X)$.

\begin{lemma}
    If $X\leq Y$ a.s.~and $\inf X +\inf Y \leq \sup Y$, then the optimization problem in (\ref{eqn:5}) is equivalent to the following optimization problem,
    \begin{align}\label{eqn:14}
         \min_{K,\tilde{Z}(\tilde{Y},\tilde{X})} & \quad \frac{(1-p)\mathbb{E}[(\tilde{X}+\tilde{Y}+\tilde{Z})^2]+pK^2}{2\Big((1-p)\mathbb{E}[\tilde{X}+\tilde{Y}+\tilde{Z}]+pK\Big)}+\mathbb{E}[Y]\nonumber\\
         \st& \quad (1-p)\mathbb{E}[\tilde{X}+\tilde{Y}+\tilde{Z}]+pK \geq \frac{1+\mathbb{P}(Y>K)}{f_{max}}
    \end{align}
where $p =\mathbb{P}(Y<K<Y+X)$, $K$ is the waiting time in state 1 and $\tilde{Z}$ is the waiting time in state 2 which is a function of the previous transmission time $\tilde{Y}$ and ACK delay $\tilde{X}$ which belongs to the set ${\{\tilde{Y}>K\}\cup \{\tilde{Y}+\tilde{X}<K\}}$.
\end{lemma}

\begin{Proof}
    From Lemma~\ref{lem:2}, $V_{i-1}+Z_{i-1} = K$. Therefore, the probability of transition from state 1 to state 1 would be also constant and is given by $p$ and hence $M\sim Geo(1-p)$. For notational convenience, represent by $\hat{Y},\hat{X}$ for any $X,Y$ that satisfy $Y<K<Y+X$ and represent by $\tilde{Y},\tilde{X}$ any $X,Y$ that do not satisfy it. Represent by $\hat{Y}'$ and $\tilde{Y}'$ independent copies of $\hat{Y}$ and $\tilde{Y}$ respectively. The expression for $\mathbb{E}\left[\int_{0}^{\tau} \Delta_t \,dt \right]$ can be evaluated for 3 cases:
    \begin{itemize}
    \item \textbf{Case 1} $M=1$
    \begin{align}\label{eqn:15}
        \mathbb{E}\left[\int_{0}^{\tau} \Delta_t \,dt |M=1 \right] &= \mathbb{E}\left[\int_{S_1}^{S_1+\tilde{Y}_1} (t-S_0) \,dt + \int_{S_1+\tilde{Y}_1}^{S_1+\tilde{Y}_1+\tilde{X}_1+\tilde{Z}} (t-S_1) \,dt\right]\nonumber\\
        &=\mathbb{E}\left[\int_{\tilde{Y}_0+\tilde{X}_0+\tilde{Z}_0}^{\tilde{Y}_0+\tilde{X}_0+\tilde{Z}_0+\tilde{Y}_1} t \,dt + \int_{\tilde{Y}_1}^{\tilde{Y}_1+\tilde{X}_1+\tilde{Z}} t \,dt\right]\nonumber\\
        &=\mathbb{E}\left[\int_{\tilde{Y}}^{\tilde{Y}+\tilde{X}+\tilde{Z}+\tilde{Y}'} t \,dt\right]
   \end{align}
   \item \textbf{Case 2} $M=2$
   \begin{align}\label{eqn:16}
        &\mathbb{E}\left[\int_{0}^{\tau} \Delta_t \,dt |M=2 \right] \nonumber\\
        &= \mathbb{E}\left[\int_{S_1}^{S_1+\hat{Y}_1} (t-S_0) \,dt +\int_{S_1+\hat{Y}_1}^{S_2+\tilde{Y}_2} (t-S_1) \,dt+\int_{S_2+\tilde{Y}_2}^{S_2+\tilde{Y}_2+\tilde{X}_2+\tilde{Z}} (t-S_2) \,dt\right]\nonumber\\      &=\mathbb{E}\left[\int_{\tilde{Y}_0+\tilde{X}_0+\tilde{Z}_0}^{\tilde{Y}_0+\tilde{X}_0+\tilde{Z}_0+\hat{Y}_1} t \,dt +\int_{\hat{Y}_1}^{K+\tilde{Y}_2} t \,dt+\int_{\tilde{Y}_2}^{\tilde{Y}_2+\tilde{X}_2+\tilde{Z}} t \,dt\right]\nonumber\\
        &=\mathbb{E}\left[\int_{\tilde{Y}}^{\tilde{Y}+\tilde{X}+\tilde{Z}+\hat{Y}} t \,dt\right]+\mathbb{E}\left[\int_{\hat{Y}}^{K+\tilde{Y}} t \,dt\right]
   \end{align}
   \item \textbf{Case 3} $M\geq 3$
   \begin{align}\label{eqn:17}
       &\mathbb{E}\left[\int_{0}^{\tau} \Delta_t \,dt |M \right] \nonumber\\
       &=\mathbb{E}\left[\int_{\tilde{Y}_0+\tilde{X}_0+\tilde{Z}_0}^{\tilde{Y}_0+\tilde{X}_0+\tilde{Z}_0+\hat{Y}_1} t \,dt + \sum_{i=1}^{M-2}\int_{\hat{Y}_i}^{K+\hat{Y}_{i+1}} t \,dt+\int_{\hat{Y}_{M-1}}^{K+\tilde{Y}_M} t \,dt+\int_{\tilde{Y}_M}^{\tilde{Y}_M+\tilde{X}_M+\tilde{Z}} t \,dt\right]\nonumber\\
       &=\mathbb{E}\left[\int_{\tilde{Y}}^{\tilde{Y}+\tilde{X}+\tilde{Z}+\hat{Y}} t \,dt\right]+(M-2)\mathbb{E}\left[\int_{\hat{Y}}^{K+\hat{Y}'} t \,dt\right]+\mathbb{E}\left[\int_{\hat{Y}}^{K+\tilde{Y}} t \,dt\right]
   \end{align}
   \end{itemize}
  
  From (\ref{eqn:15}), (\ref{eqn:16}) and (\ref{eqn:17}) we can find $\mathbb{E}\left[\int_{0}^{\tau} \Delta_t \,dt \right]$ as follows,
  \begin{align}\label{eqn:18}
      &\mathbb{E}\left[\int_{0}^{\tau} \Delta_t \,dt \right]\nonumber\\
      &=\sum_{m=1}^{\infty}\mathbb{P}(M=m)\mathbb{E}\left[\int_{0}^{\tau} \Delta_t \,dt|M=m \right]\nonumber\\
      &=(1-p)\mathbb{E}\left[\int_{\tilde{Y}}^{\tilde{Y}+\tilde{X}+\tilde{Z}+\tilde{Y}'} t \,dt\right]+(1-p)p\left(\mathbb{E}\left[\int_{\tilde{Y}}^{\tilde{Y}+\tilde{X}+\tilde{Z}+\hat{Y}} t \,dt\right]+\mathbb{E}\left[\int_{\hat{Y}}^{K+\tilde{Y}} t \,dt\right]\right)\nonumber\\
      &\quad +\sum_{m=3}^{\infty}(1-p)p^{m-1}\left(\mathbb{E}\left[\int_{\tilde{Y}}^{\tilde{Y}+\tilde{X}+\tilde{Z}+\hat{Y}} t \,dt\right]+(m-2)\mathbb{E}\left[\int_{\hat{Y}}^{K+\hat{Y}'} t \,dt\right]+\mathbb{E}\left[\int_{\hat{Y}}^{K+\tilde{Y}} t \,dt\right]\right)\nonumber\\
      &=(1-p)\mathbb{E}\left[\int_{\tilde{Y}}^{\tilde{Y}+\tilde{X}+\tilde{Z}+\tilde{Y}'} t \,dt\right]+p\mathbb{E}\left[\int_{\tilde{Y}}^{\tilde{Y}+\tilde{X}+\tilde{Z}+\hat{Y}} t \,dt\right]+p\mathbb{E}\left[\int_{\hat{Y}}^{K+\tilde{Y}} t \,dt\right]\nonumber\\
      &\quad +\mathbb{E}\left[\int_{\hat{Y}}^{K+\hat{Y}'} t \,dt\right]\sum_{m=3}^{\infty}(1-p)p^{m-1}(m-2)\nonumber\\
      &=\frac{1}{2}\Bigg(\mathbb{E}\left[(\tilde{Y}+\tilde{X}+\tilde{Z})^2\right]+2\mathbb{E}\left[\tilde{Y}+\tilde{X}+\tilde{Z}\right]\left(p\mathbb{E}[\hat{Y}]+(1-p)\mathbb{E}[\tilde{Y}]\right)\Bigg) \nonumber\\
      & \quad +\frac{1}{2}\Bigg(p\left(K^2+2K\mathbb{E}[\tilde{Y}]\right)+\frac{p^2}{1-p}\left(K^2+2K\mathbb{E}[\hat{Y}]\right)\Bigg)\nonumber\\
      &=\frac{(1-p)\Big(\mathbb{E}\left[(\tilde{Y}+\tilde{X}+\tilde{Z})^2\right]+2\mathbb{E}\left[\tilde{Y}+\tilde{X}+\tilde{Z}\right]\mathbb{E}[Y]\Big)+p\left(K^2+2K\mathbb{E}[Y]\right)}{2(1-p)}
  \end{align}
  Similarly, $\mathbb{E}[\tau]$ can be found as follows,
  \begin{align}\label{eqn:19}
      \mathbb{E}[\tau] &= \sum_{m=1}^{\infty}\mathbb{P}(M=m)\mathbb{E}[\tau|M=m]\nonumber\\
      &=(1-p)\mathbb{E}[\tilde{Y}+\tilde{X}+\tilde{Z}]+\sum_{m=2}^{\infty}(1-p)p^{m-1}\Big(K(m-1)+\mathbb{E}[\tilde{Y}+\tilde{X}+\tilde{Z}]\Big)\nonumber\\
      &=\mathbb{E}[\tilde{Y}+\tilde{X}+\tilde{Z}]+K\sum_{m=2}^{\infty}(1-p)p^{m-1}(m-1)\nonumber\\
      &=\frac{(1-p)\mathbb{E}[\tilde{Y}+\tilde{X}+\tilde{Z}]+pK}{1-p}
  \end{align}
  and $\mathbb{E}[M+\textbf{1}_e]$ can be found as follows,
  \begin{align}\label{eqn:20}
      \mathbb{E}\left[M+\textbf{1}_e\right]&=\mathbb{E}[M]+\mathbb{E}\left[\textbf{1}_{\{\tilde{Y}>K\}}\right]\nonumber\\
      &=\frac{1}{1-p}+ \mathbb{P}(Y>K|\{Y>K\}\cup  \{Y+X<K\})\nonumber\\
      &=\frac{1}{1-p}+ \frac{\mathbb{P}(Y>K)}{\mathbb{P}(\{Y>K\}\cup  \{Y+X<K\})}\nonumber\\
      &=\frac{1+\mathbb{P}(Y>K)}{1-p}
  \end{align}
Substituting (\ref{eqn:18}), (\ref{eqn:19}) and (\ref{eqn:20}) in problem (\ref{eq:2}) yields the required result.
\end{Proof}

\begin{theorem}\label{thrm:1}
    If $\mathbb{E}[Y^2]<\infty$, then the optimal policy that minimizes (\ref{eqn:14}) achieves a lower average AoI than any optimal policy that always waits for ACKs before taking the next sample.
\end{theorem}

\begin{Proof}
    If $K \to \infty$, then $p \to 0$, $\mathbb{P}(Y>K) \to 0$, $\mathbb{P}(Y+X<K) \to 1$, $\tilde{X} \to X$ and $\tilde{Y} \to Y$. Let $F_{Y+X}$ denote the distribution function of $Y+X$. Since $p<\mathbb{P}(K<Y+X)$, it can be seen that $pK^2\leq \int_{K}K^2\,dF_{Y+X} \leq \int_{K}(y+x)^2\,dF_{Y+X}$. If $\mathbb{E}[Y^2]<\infty$, then  $\mathbb{E}[(Y+X)^2]<\infty$. Therefore, $\int_{K}(y+x)^2\,dF_{Y+X} \to 0$ as $K$ increases. Hence, $pK^2 \to 0$ for large $K$. Additionally, $pK^2 \to 0$ implies $pK \to 0$. Therefore, as $K \to \infty$, the optimization problem in (\ref{eqn:14}) will reduce to the following,
    \begin{align} \label{eqn:21}
        \min_{\tilde{Z}(Y,X)} & \quad \frac{\mathbb{E}\left[(X+Y+\tilde{Z})^2\right]}{2\mathbb{E}\left[X+Y+\tilde{Z}\right]}+\mathbb{E}[Y]\nonumber\\
         \st& \quad \mathbb{E}\left[X+Y+\tilde{Z}\right] \geq \frac{1}{f_{max}}
    \end{align}
Problem in (\ref{eqn:21}) is the exact optimization problem that we must solve for an optimal policy which always waits for an ACK to sample the next value \cite{twd2020,urtwd2022}. Let $\alpha(K) $ be the optimal value of problem (\ref{eqn:14}) for a given $K$. Then optimal value of (\ref{eqn:14}) is simply $\inf_{K}\alpha(K)$ and the optimal solution of (\ref{eqn:21}) is $\lim_{K \to \infty} \alpha(K)$. This proves the required result.
\end{Proof}

Theorem~\ref{thrm:1} shows that the optimal policy that solves (\ref{eqn:14}) always outperforms any optimal policy constructed which always waits for the ACK of the previous sample before sampling the next. Next, we solve for the optimal functional $\tilde{Z}$ for a fixed $K$. 

\begin{theorem}
    For a fixed  $K$, the optimal functional $\tilde{Z}$ that solves (\ref{eqn:14}) is given by,
    \begin{align}
        \tilde{Z} = \Big(\beta - (\tilde{X}+\tilde{Y})\Big)^+
    \end{align}
    where $\beta >0$ and satisfies,
    \begin{align} \label{eqn:23}
         (1-p)\mathbb{E}&\left[\max(\beta,\tilde{X}+\tilde{Y})\right]+pK \nonumber\\
         &=\max\Bigg(\frac{1+\mathbb{P}(Y>K)}{f_{max}},\frac{(1-p)\mathbb{E}\left[\max(\beta^2,(\tilde{X}+\tilde{Y})^2)\right]+pK^2}{2\beta}\Bigg) 
    \end{align}
\end{theorem}

\begin{Proof}
    For a fixed $K$ the optimization problem is similar to the problem in \cite{ys17}. Therefore, we follow similar arguments and techniques. We use the one-sided G\^{a}teaux derivative to solve for the optimal functional. Let $U =\tilde{X}+\tilde{Y}$ and its distribution function be $F_u$. Let $T=\frac{1+\mathbb{P}(Y>K)}{f_{max}}$. Then, for a fixed $K$, the  Lagrangian of (\ref{eqn:14}) is as follows,
    \begin{align}
        L(\tilde{Z},\gamma,\lambda) = \frac{(1-p)\mathbb{E}\left[(U+\tilde{Z})^2\right]+pK^2}{2\Big((1-p)\mathbb{E}\left[U+\tilde{Z}\right]+pK\Big)} -\int\gamma(u)\tilde{Z}\, dF_u+\lambda\Big(T-(1-p)\mathbb{E}\left[U+\tilde{Z}\right]-pK\Big) \label{eqn:24}
    \end{align}
    Where $\gamma(u)$ and $\lambda$ are the Lagrangian multipliers corresponding to the non-negativity of $\tilde{Z}$ and the sampling constraint respectively . The one-sided G\^{a}teaux derivative $\delta_{w}$ of the Lagrangian for an arbitrary functional $w$ is,
    \begin{align}
        \delta_{w}=\lim_{\epsilon \to 0} \pdv{L(\tilde{Z}+\epsilon w,\gamma,\lambda)}{\epsilon}
    \end{align}
    Let $Q=(1-p)\mathbb{E}\left[(U+\tilde{Z})^2\right]+pK^2$ and $R=(1-p)\mathbb{E}\left[U+\tilde{Z}\right]+pK$. Then, $\delta_{w}$ can be evaluated as follow,
    \begin{align}
        \delta_{w}=& \lim_{\epsilon \to 0}\frac{R \int 2(1-p)(u+\tilde{z}+\epsilon w)w\,dF_{u} -(1-p)Q\int w\, dF}{2R^2}\nonumber\\
        &-\int \gamma(u)w\,dF_u - \lambda(1-p)\int w \,dF_u \nonumber\\
        =&\int \Bigg[\frac{(1-p)\left(2R(u+\tilde{z})-Q\right)}{2R^2} -\gamma(u) -\lambda(1-p)\Bigg]w\,dF_u \label{eqn:26}
    \end{align}
    For the optimal functional $\tilde{Z}$, $\delta_w  \geq 0, \ \forall w$. Since $\delta_w=-\delta_{-w}$,  for the optimal functional $\delta_w = 0, \ \forall w$. Since $w$ is an arbitrary function, the optimal functional should satisfy the following,
    \begin{align}
        u+\tilde{z} &=  \frac{Q}{2R}+\lambda R + \frac{\gamma(u) R}{1-p} \label{eqn:27}\\
        \gamma(u)\tilde{z} &= 0 \label{eqn:28}\\ 
        \lambda(T-R) &= 0 \label{eqn:29}
    \end{align}
    where (\ref{eqn:28}) and (\ref{eqn:29}) are from complementary slackness conditions. If $\gamma(u)>0$, then $\tilde{z} = 0$. If $\gamma = 0$, then  $u+\tilde{z}= \beta$ where $\beta = \frac{Q}{2R}+\lambda R$. Therefore, the optimal functional is a threshold policy. To find the optimal $\beta$, note that $R$ is an increasing function of threshold $\beta$ and $\frac{Q}{2R}$ is exactly the term we are optimizing. Thus, if we can find $\beta$ such that $R = T$ and  that particular $\beta $ satisfies $\beta \geq \frac{Q}{2R}$, then that $\beta$ along with $\lambda = (\beta-\frac{Q}{2R})/R$ satisfies the optimality conditions given (\ref{eqn:27}), (\ref{eqn:28}) and (\ref{eqn:29}). However, if at that point $\beta < \frac{Q}{2R}$, then we need to increase $\beta$ till $\beta = \frac{Q}{2R}$. Validity of this second criterion is guaranteed by simply noting that the derivative of $\frac{Q}{2R}$ with respect to $\beta$ is negative when $\beta < \frac{Q}{2R}$ (see Appendix~\ref{apd:converge} for a proof). Therefore, as $\beta$ increases, $\frac{Q}{2R}$ decreases. Since we have increased $\beta$ beyond the point where $R = T$, $\lambda$ must be zero and the optimality conditions are again achieved  at $\beta = \frac{Q}{2R}$. Combining all these together yields (\ref{eqn:23}).
\end{Proof}

\begin{figure}[t]
    \centering
    \begin{subfigure}[b]{0.3\textwidth}
         \centering
         \includegraphics[width=\textwidth]{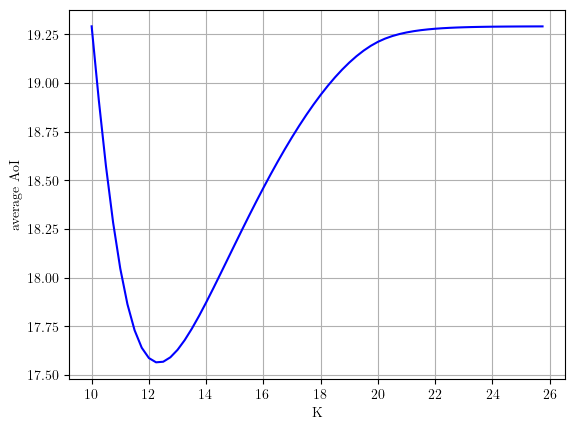}
         \caption{$1/f_{max}=8$}
     \end{subfigure}
    \begin{subfigure}[b]{0.3\textwidth}
         \centering
         \includegraphics[width=\textwidth]{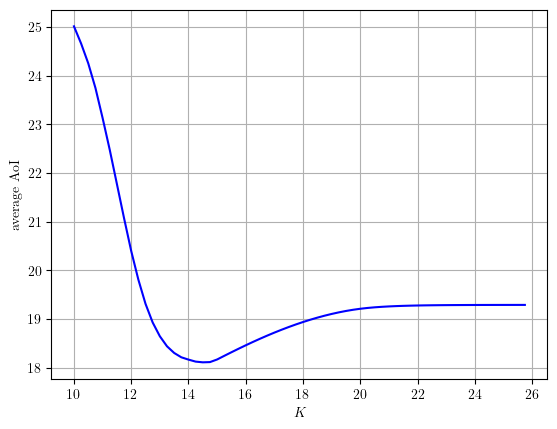}
         \caption{$1/f_{max}=14$}
     \end{subfigure}
     \begin{subfigure}[b]{0.3\textwidth}
         \centering
         \includegraphics[width=\textwidth]{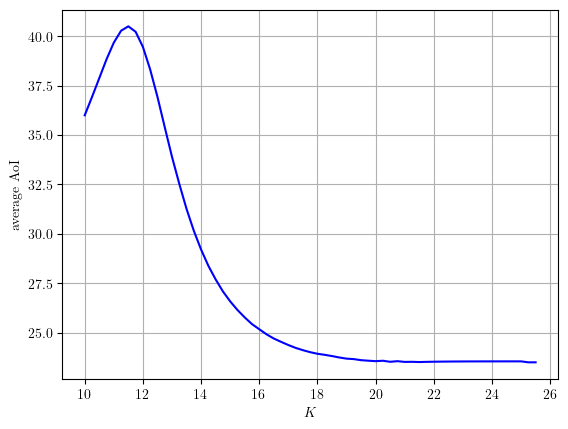}
         \caption{$1/f_{max}=25$}
     \end{subfigure}
    \caption{Variation of average AoI with $K$ for $Y\sim (10+Exp(1))$ and $X\sim Uniform[0,10]$.}
    \label{fig:KvAoI}
\end{figure}

Similar to \cite{ys17}, we can use a bisection method to solve for the optimal $\beta$ for a given $K$. However, the optimization with respect to $K$ is a non-convex problem (see Fig.~\ref{fig:KvAoI}). Algorithm~\ref{algo:1} below provides a sub-optimal descent type algorithm to search for $K$. 

\begin{algorithm} [t]
    \caption{Algorithm for finding optimal $K$ and $\beta$}\label{algo:1}
    \begin{algorithmic}
        \REQUIRE $K=\inf Y$, $l_0=0$, $\{u_0,old,new,K_0\}$ sufficiently large, $\{\lambda,\epsilon\}$ sufficiently small
        $K^* = K_0$, $\beta^*=\beta_{K_0}$ (optimal $\beta$ for $K=K_0)$, $old>new$
        \WHILE{($old >new$ \AND $K<K_0$)}
            \item $old=new$, $u=u_0$, $l=l_0$
            \WHILE{$(u-l)>\epsilon$}
                \item $B_K=\{Y<K<Y+X\}$
                \item $\beta_K =\frac{u+l}{2}$,\quad$p=\mathbb{P}(B_K)$
                \item $Q=(1-p)\mathbb{E}\left[\max\{\beta_K^2,(X+Y)^2\}|\overline{B_K}\right]+pK^2$
                \item $R =(1-p)\mathbb{E}\left[\max\{\beta_K,(X+Y)\}|\overline{B_K}\right]+pK$
                \item $T=\frac{1+\mathbb{P}(Y>K)}{f_{max}}$
                \item $diff = R-\max\{T,\frac{Q}{2\beta_K}\}$
                \IF{$diff\leq 0$}
                    \item $l=\beta_K$
                \ELSE 
                    \item $u=\beta_K$
                \ENDIF
            \ENDWHILE
            \item $new = \frac{Q}{2R}$
            \IF{$old-new>0$}
                \item $K^* = K$, $\beta^*=\beta_K$
            \ENDIF
            \item $K=K+\lambda$
        \ENDWHILE
    \end{algorithmic}
\end{algorithm}

\begin{theorem}\label{thrm:3}
    If there is a $K$ such that $\max\left\{\sup Y,\frac{1}{f_{max}}\right\} <K<\inf Y + \inf X$, then there exists a periodic sampling policy which always has a lower average AoI than any optimal policy constructed where one always waits for an ACK before sampling the next. The period of the optimal periodic sampling policy is the smallest $K$ that satisfies the above inequality. 
\end{theorem}

\begin{Proof}
    Note that if $\sup Y <K<\inf Y + \inf X$, then $p=1$. Therefore, only state 1 to state 1 transitions would be taken and the average AoI can be evaluated using only one of the state 1 to state 1 transition. Let $AoI_{per}(K)$ denote the average AoI for a periodic sampling policy with period $K$. Then $AoI_{per}(K)$ is given by,
    \begin{align}
        AoI_{per}(K) = \frac{\mathbb{E}\left[\int_Y^{K+Y'}t\,dt\right]}{K} =\frac{K^2+2K\mathbb{E}\left[Y\right]}{2K}= \frac{K}{2}+\mathbb{E}\left[Y\right]
    \end{align}
  Now, consider the optimal value of problem (\ref{eqn:21}), 
    \begin{align}
        \frac{\mathbb{E}\left[(X+Y+\tilde{Z})^2\right]}{2\mathbb{E}\left[X+Y+\tilde{Z}\right]}+\mathbb{E}\left[Y\right] &\geq \frac{\left(\mathbb{E}\left[X+Y+\tilde{Z}\right]\right)^2}{2\mathbb{E}\left[X+Y+\tilde{Z}\right]}+\mathbb{E}\left[Y\right]\nonumber \\
        &=\frac{\mathbb{E}\left[X+Y+\tilde{Z}\right]}{2}+\mathbb{E}\left[Y\right]\nonumber \\
        &\geq \frac{\inf Y+\inf X}{2}+\mathbb{E}\left[Y\right] \nonumber \\
        &>\frac{K}{2}+\mathbb{E}\left[Y\right]
    \end{align}
    Therefore, if $K$ further satisfies $K>\frac{1}{f_{max}}$, then the periodic sampling policy is always better than any policy constructed by waiting for ACKs always.
\end{Proof}

\section{Numerical Results}
In this section, we compare the performance of our optimal policy with the optimal policy \textit{wait-for-ACK} obtained by solving the problem (\ref{eqn:21}) and a periodic sampling policy \textit{periodic+preempt}  that satisfies the sampling constraint exactly. In the periodic sampling policy, we always enable preemptive transmissions if an ACK indicates that the channel is serving a corrupted sample. We compare the results under different distributions for $Y$ and $X$. 

In the first experiment, we take the distribution of $Y$ to be a shifted exponential (i.e., $Y= C + \bar{Y}$ where $\bar{Y}\sim exp(\gamma)$ and $C>0$) and we take the distribution of $X$ to be uniform in the interval from zero to $\inf Y$ (i.e., $X \sim Unif[0,C]$). In the second experiment, the distribution of $Y$ is again chosen to be the same shifted exponential as before but here we take $X$ to take the constant value $\frac{C}{2}$. In the third experiment, we set $Y$ to be a constant ($C$) and set $X$ to have the same uniform distribution as in first experiment. To compare the policies, we plot the variation of the average AoI with $\frac{1}{f_{max}}$.  

\begin{figure}[t]
    \centering
     \begin{subfigure}[b]{0.49\textwidth}
         \centering
         \includegraphics[width=\textwidth]{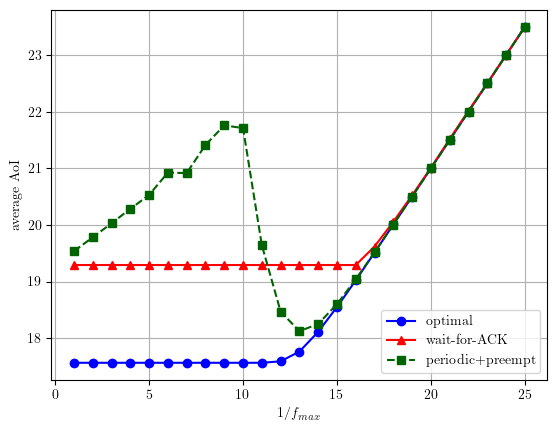}
         \caption{$C=10$, $\gamma = 1$}
     \end{subfigure}
    \begin{subfigure}[b]{0.49\textwidth}
         \centering
         \includegraphics[width=\textwidth]{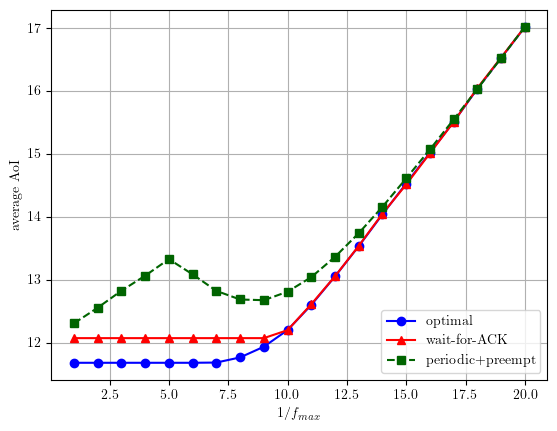}
         \caption{$C=5$, $\gamma = 0.5$}
     \end{subfigure}
     \begin{subfigure}[b]{0.49\textwidth}
         \centering
         \includegraphics[width=\textwidth]{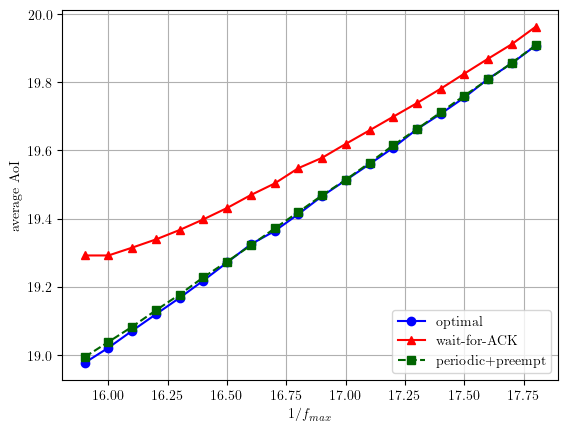}
         \caption{$C=10$, $\gamma = 1$,  zoomed low $f_{max}$ region}
     \end{subfigure}
    \begin{subfigure}[b]{0.49\textwidth}
         \centering
         \includegraphics[width=\textwidth]{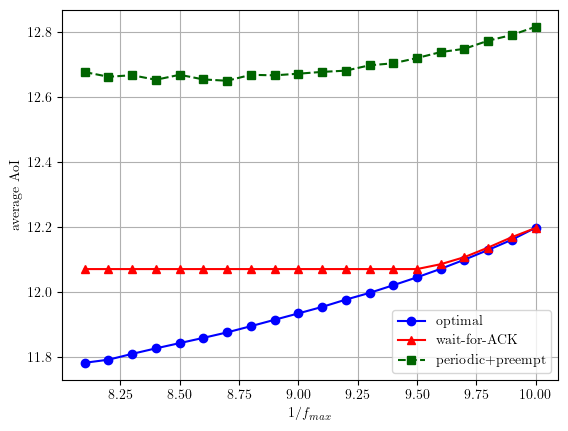}
         \caption{$C=5$, $\gamma = 0.5$, zoomed low $f_{max}$ region}
     \end{subfigure}
    \caption{Variation of the average AoI with maximum allowable sampling rate for $Y\sim (C+Exp(\gamma))$ and $X\sim Uniform[0,C]$.}
    \label{fig:exp1}
\end{figure}

\begin{figure}[t]
    \centering
    \begin{subfigure}[b]{0.49\textwidth}
         \centering
         \includegraphics[width=\textwidth]{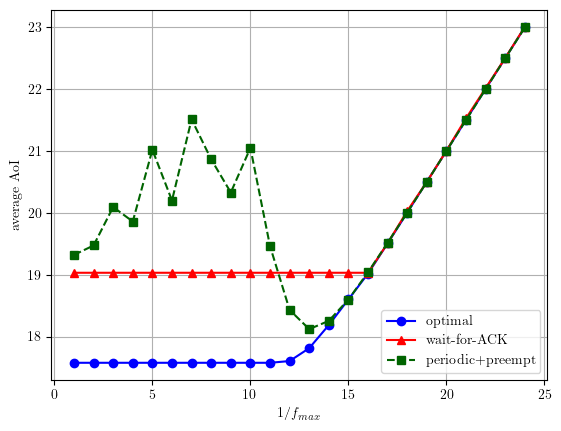}
         \caption{$C=10$, $\gamma = 1$}
     \end{subfigure}
    \begin{subfigure}[b]{0.49\textwidth}
         \centering
         \includegraphics[width=\textwidth]{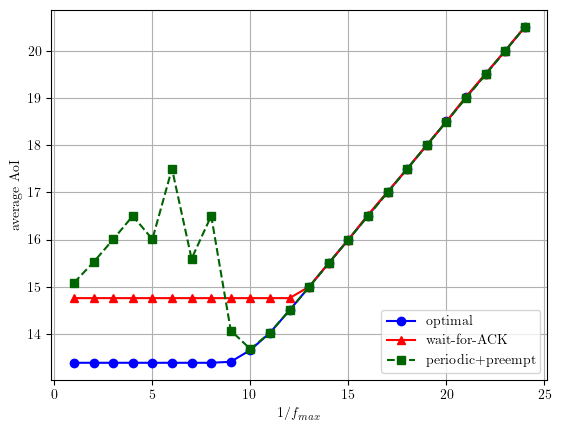}
         \caption{$C=8$, $\gamma = 2$}
     \end{subfigure}
    \caption{Variation of the average AoI with maximum allowable sampling rate for $Y\sim (C+Exp(\gamma))$ and $X=C/2$.}
    \label{fig:exp2}
\end{figure}

\begin{figure}
    \centering
    \includegraphics[width=0.49\textwidth]{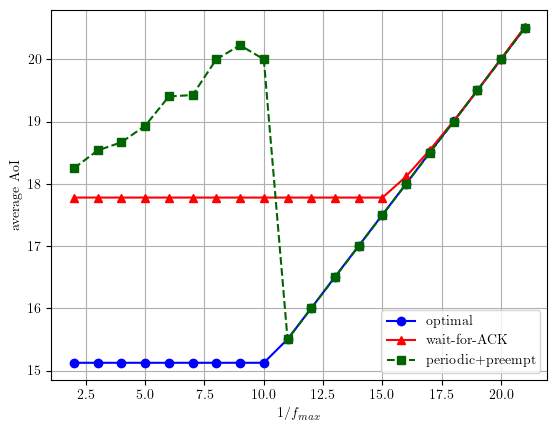}
    \caption{Variation of the average AoI with maximum allowable sampling rate for $Y=10 $ and $X\sim Uniform[0,10]$.}
    \label{fig:exp3}
\end{figure}

As seen in Figs.~\ref{fig:exp1}, \ref{fig:exp2} and \ref{fig:exp3}, for higher values of $f_{max}$ (lower values of $\frac{1}{f_{max}}$), our policy is significantly better than the other two policies considered. However, for lower values of $f_{max}$ the average AoI of all three policies tend to be similar. This is because, even though our policy allows sampling at a faster rate when in state 1 as we can compensate by waiting longer in state 2, if the rate of sampling in state 1 is too fast, corrupted samples will arise more frequently and as a result the required waiting time to satisfy the sampling constraint for low values of $f_{max}$ will be much larger. Additionally, long cycles of state 1 to state 1 transitions will be less often in this case. Therefore, the optimal value $K$ to sample in state 1 would generally increase with $\frac{1}{f_{max}}$. As $K$ increases, corrupted samples will be less frequent and ACKs would arrive before $K$ time units have elapsed more often. Hence, the similarity in the three curves for lower values of $f_{max}$.

Fig.~\ref{fig:exp1} shows that when the variation of $Y$ is greater, then the periodic sampling policy is far from optimal, however when the values of $Y$  become concentrated at its lower bound ($\gamma=1$ or $Y=C$), the periodic sampling policy closely follows the optimal policy when $1/f_{max} > \inf Y$. However, at any given value of $1/f_{max}$, the periodic sampling policy never goes below the curve of the optimal policy. This indicates that even in the absence of the sampling constraint, a periodic sampling policy with any period (i.e., sampling at a rate other than $f_{max}$) will not be better than the optimal policy constructed here. As seen by the presented figures, our simulation results validate our theoretical development of the optimal policy for the given system model.

\section{Conclusion}
In this work, we have introduced a new system model which facilities early sampling and transmission before receiving an ACK. We have shown through theoretical results and simulations that it is not always optimal to wait for ACKs before sampling when there is a delay in the feedback channel. The system model introduced here may be an optimistic abstraction of what is really happening in the real world scenarios when collisions in transmissions occur (here we assumed that the already transmitting packet is not affected but the new arriving packet is corrupted; in real world scenarios both packets may be corrupted or none may be corrupted and the new arriving packet may be queued up). Our work could provide a useful first perspective when tackling more complex scenarios. Future directions of work may include considering that both the transmitting and the new sample get corrupted in case of a collision, samples obtained before an ACK are not corrupted but just queued up to be transmitted in the forward channel, and corrupted samples are transmitted without any preemptive transmissions.

\begin{appendices}
\section{Proof of Convexity of (\ref{eqn:5}) and (\ref{eqn:6})}\label{apd:cvx} 
  
  Here, we show that for a given policy in $Z$, the problem is convex with respect to the function $\tilde{Z}$. Given the policy in $Z$, the distribution of $M$, $Y_M$ and $X_M$ are fixed. Let $F$ denote the joint distribution of $M$, $(Y_i,X_i)_{i=1}^M$ and $Y_{M+1}$. Let us define,
  \begin{align}
       f(\tilde{Z})&= \mathbb{E}\left[\sum_{i=1}^{M-1} \int_{Y_i}^{V_{i-1}+Z_{i-1}+Y_{i+1}}t \,dt\right] +\mathbb{E}\left[\int_{Y_M}^{Y_M+X_M+\tilde{Z}+Y_{M+1}}t \,dt  \right] \\
       g(\tilde{Z}) &= \mathbb{E}\left[\sum_{i=1}^{M-1}(V_{i-1}+Z_{i-1})+Y_M+X_M+\tilde{Z}\right]
  \end{align}
  Then, the functions of interest will be given by $h_1(\tilde{Z})=f(\tilde{Z})/g(\tilde{Z})$ and $h_2(\tilde{Z})=f(\tilde{Z})-cg(\tilde{Z})$. Let $w$ be an arbitrary functional in the functional space of $\tilde{Z}$. Then, the functional $h_i(\tilde{Z})$ is said to be convex with respect to $\tilde{Z}$ iff $\frac{d^2h_i(\tilde{Z}+\epsilon w)}{d\epsilon^2}\geq0, \ \forall w$ where $i=1,2$. For notional convenience let the functionals $h_i(\tilde{Z}+\epsilon w)$, $f(\tilde{Z}+\epsilon w)$, $g(\tilde{Z}+\epsilon w)$ be represented as $h_i$, $f$, $g$. Then,
  \begin{align}
      \dv{f}{\epsilon} &=\frac{d}{d\epsilon} \int \left(\sum_{i=1}^{M-1}\left[\frac{(V_{i-1}+Z_{i-1}+Y_{i+1})^2-Y_i^2}{2}\right]+\frac{(Y_M+X_M+\tilde{Z}+\epsilon w+Y_{M+1})^2-Y_M^2}{2} \right)\,dF \nonumber\\
      &= \int \frac{d}{d\epsilon}\left(\sum_{i=1}^{M-1}\left[\frac{(V_{i-1}+Z_{i-1}+Y_{i+1})^2-Y_i^2}{2}\right]+\frac{(Y_M+X_M+\tilde{Z}+\epsilon w+Y_{M+1})^2-Y_M^2}{2} \right)\,dF \nonumber\\
      &=\int (Y_M+X_M+\tilde{Z}+\epsilon w+Y_{M+1})w\,dF \label{eqn:df} \\
      \dv{g}{\epsilon} &= \frac{d}{d\epsilon} \int \left(\sum_{i=1}^{M-1}(V_{i-1}+Z_{i-1})+Y_M+X_M+\tilde{Z}+\epsilon w\right)\,dF  = \int w\,dF \label{eqn:dg}
  \end{align}
  The interchange of the integral and the derivative in (\ref{eqn:df}) and (\ref{eqn:dg}) can be justified similar to \cite[Lem.~2]{ys17}. In addition,
  \begin{align}
      \dv{h_1}{\epsilon}&= \frac{\int (Y_M+X_M+\tilde{Z}+\epsilon w+Y_{M+1})w\,dF}{g} - \frac{f\int w\,dF}{g^2}\\
      \frac{d^2h_1}{d\epsilon}&=\frac{\int w^2\,dF }{g}-\frac{2}{g^2}\int (Y_M+X_M+\tilde{Z}+\epsilon w+Y_{M+1})w\,dF \int w\,dF+\frac{2f\left(\int w\,dF\right)^2}{g^3}\nonumber\\
      &=\frac{\left(\int w\,dF\right)^2}{g^3}\int \Bigg(\frac{g^2w^2}{\left(\int w\,dF\right)^2}-\frac{2gw}{\int w\,dF}(Y_M+X_M+\tilde{Z}+\epsilon w+Y_{M+1})-Y_M^2\nonumber\\
      &\qquad \qquad+(Y_M+X_M+\tilde{Z}+\epsilon w+Y_{M+1})^2+ \sum_{i=1}^{M-1}\left[(V_{i-1}+Z_{i-1}+Y_{i+1})^2-Y_i^2\right]\Bigg)\,dF\nonumber\\
      &=\frac{\left(\int w\,dF\right)^2}{g^3}\int \Bigg( \Big(\frac{gw}{\int w\,dF}-(Y_M+X_M+\tilde{Z}+\epsilon w+Y_{M+1})\Big)^2 +Z_0^2-Y_1^2+2Z_0Y_2\nonumber\\
      &\qquad \qquad + \sum_{i=2}^{M-1}\left[(V_{i-1}+Z_{i-1})^2+2Y_{i+1}(V_{i-1}+Z_{i-1})\right]\Bigg)\,dF
  \end{align}
  Since $Z_0\geq Y_1$ and $V_i$, $Z_i$, $Y_i$ are all non-negative, $\frac{d^2h_1}{d\epsilon}\geq 0, \ \forall w$ in the functional of $\tilde{Z}$. Therefore, $h_1(\tilde{Z})$ is convex with respect to the functional $\tilde{Z}$. In the same manner,
  \begin{align}
      \dv{h_2}{\epsilon}&=\int (Y_M+X_M+\tilde{Z}+\epsilon w+Y_{M+1}-c)w\,dF\\
      \frac{d^2h_2}{d\epsilon}&= \int w^2\,dF \geq 0
  \end{align}
  Therefore, $h_2(\tilde{Z})$ is convex with respect to the functional $\tilde{Z}$.
  \section{Proof of Convergence of (\ref{eqn:23})}\label{apd:converge}
  Let $Q$, $R$, $U$ and $F_u$ follow the same definitions as in (\ref{eqn:24}). Let $\beta$ be value of the threshold employed in state 2 and $f_u$ be the pdf of $U$. Then,
  \begin{align}
      Q&=(1-p)\mathbb{E}\left[\max(\beta^2,(\tilde{X}+\tilde{Y})^2)\right]+pK^2\nonumber\\
      &=(1-p)\left(\beta^2\int_0^\beta\,dF_u+\int_\beta u^2\,dF_u\right)+pK^2\nonumber\\
      &=(1-p)\left(\beta^2\int_0^\beta\,dF_u+\mathbb{E}\left[U^2\right]-\int_0^\beta u^2\,dF_u\right)+pK^2\\
      \dv{Q}{\beta}&=(1-p)\left(2\beta\int_0^\beta\,dF_u+\beta^2f_u(\beta)-\beta^2f_u(\beta)\right)\nonumber\\
      &=(1-p)\left(2\beta\int_0^\beta\,dF_u\right)\\      
      R&=(1-p)\mathbb{E}\left[\max(\beta,\tilde{X}+\tilde{Y})\right]+pK\nonumber\\
      &=(1-p)\left(\beta\int_0^\beta\,dF_u+\int_\beta u\,dF_u\right)+pK\nonumber\\
      &=(1-p)\left(\beta\int_0^\beta\,dF_u+\mathbb{E}\left[U\right]-\int_0^\beta u\,dF_u\right)+pK\\
      \dv{R}{\beta}&=(1-p)\left(\int_0^\beta\,dF_u+\beta f_u(\beta)-\beta f_u(\beta)\right)\nonumber\\
      &=(1-p)\int_0^\beta\,dF_u\\
      \dv{(\frac{Q}{R})}{\beta} &= \frac{(1-p)\left(2\beta\int_0^\beta\,dF_u\right)R-(1-p)\left(\int_0^\beta\,dF_u\right)Q}{R^2}\nonumber\\
      &=\frac{2(1-p)\left(\int_0^\beta\,dF_u\right)R\left(\beta-\frac{Q}{2R}\right)}{R^2}
  \end{align}
 If $\frac{Q}{2R}>\beta$ at any given $\beta$, then $\dv{(\frac{Q}{R})}{\beta}<0$. Therefore, increasing $\beta$ would decrease $\frac{Q}{2R}$. Thus, the convergence of (\ref{eqn:23}) is guaranteed by starting from $\beta$ such that $\frac{Q}{2R}>\beta$ and then increasing $\beta$ until $\frac{Q}{2R}=\beta$.
\end{appendices}

\bibliographystyle{unsrt}
\bibliography{refs}

\end{document}